\documentclass[]{mn2e}
\usepackage{graphicx}
\usepackage{epsfig}
\newcommand       \Angstrom     {\,{\rm \AA}}          

\newcommand       \cm           {\,{\rm cm}}

\newcommand	\g	       {\,{\rm g}}

\newcommand       \mum          {\,{\rm \mu m}}

\newcommand	  \ppm		{\,{\rm ppm}}

\newcommand	  \cabs		{C_{\rm abs}}
\newcommand	  \cpol		{C_{\rm pol}}
\newcommand	  \cpar		{C_{\rm abs}^{\|}}
\newcommand	  \cper		{C_{\rm abs}^{\bot}} 
\newcommand	  \copol	{C_{\rm pol}^{\rm obl}}
\newcommand	  \cppol	{C_{\rm pol}^{\rm pro}}

\newcommand	  \tauor	{\tau_{3.4}}
\newcommand	  \tausil	{\tau_{9.7}}
\newcommand	  \vsil	        {V_{\rm sil}}
\newcommand	  \vor	        {V_{\rm carb}}
\newcommand	  \Asil	        {A_{\rm 9.7}}
\newcommand	  \Ach	        {A_{\rm 3.4}}
\newcommand	  \Psil	        {P_{\rm 9.7}}
\newcommand	  \Pch	        {P_{\rm 3.4}}
\newcommand \PTOA {\left(P_{3.4}/A_{3.4}\right)/\left(P_{9.7}/A_{9.7}\right)}

\newcommand	  \csun         {\left[{\rm C/H}\right]_{\odot}}
\newcommand	  \cism         {\left[{\rm C/H}\right]_{\rm ISM}}
\newcommand	  \cdust        {\left[{\rm C/H}\right]_{\rm dust}}
\newcommand	  \cgas        {\left[{\rm C/H}\right]_{\rm gas}}
\newcommand       \cBstar       {\left[{\rm C/H}\right]_{\star}}
\newcommand	  \cpah        {\left[{\rm C/H}\right]_{\rm PAH}}
\newcommand       \mnras        {MNRAS}
\newcommand       \apj          {ApJ}
\newcommand       \aj           {AJ}
\newcommand       \aanda        {A\&A}
\newcommand       \araa         {ARA\&A}
\newcommand       \apjs         {ApJS}
\newcommand       \qjras        {QJRAS}
\title{Spectropolarimetric Constraints on
       the Nature of Interstellar Grains}

\author[Li, Liang, \& Li]
       {Qi Li$^{1,2}$,
        S.L. Liang$^{2}$, and
        Aigen Li$^{2}$\thanks{%
                E-mail: lia@missouri.edu}\\
        $^1$Department of Physics, Xiangtan University,
            Xiangtan 411105, Hunan Province, China\\
        $^2$Department of Physics and Astronomy,
             University of Missouri,
             Columbia, MO 65211, USA
             }
\begin{document}
\date{Received date  / Accepted date }
\pagerange{\pageref{firstpage}--\pageref{lastpage}} \pubyear{2013}

\maketitle

\label{firstpage}
\begin{abstract}
While it is well recognized that interstellar grains 
are made of amorphous silicates
and some form of carbonaceous materials,
it remains debated regarding what exact chemical 
and physical form the  carbonaceous component takes. 
Contemporary grain models assume that the silicate 
and carbon components are either physically separated,
or they form a core-mantle structure, 
or they agglomerate to form porous composites. 
The core-mantle model posits that the mantle is
made of some sort of aliphatic hydrocarbon materials 
and is responsible for the 3.4$\mum$ absorption feature 
ubiquitously seen in the diffuse interstellar medium (ISM) 
of the Milky Way and external galaxies. 
This model is challenged 
by the nondetection of polarization 
in the 3.4$\mum$ absorption feature
as the 9.7$\mum$ silicate feature 
is observed to be polarized.
%
To alleviate this challenge, we calculate 
the degree of polarization 
of the 3.4$\mum$ 
feature for 
{\it spheroidal} silicate dust coated by a layer
of {\it spherical} aliphatic hydrocarbon.
It is found that the 3.4$\mum$ feature polarization
still exceeds the observational upper limit,
even though {\it spherical} aliphatic hydrocarbon
mantles are expected to cause much less polarization
than {\it nonspherical} (e.g., spheroidal) mantles.
%
We have also shown that the composite grain model
which consists of amorphous silicate, 
aliphatic hydrocarbon, and vacuum
also predicts the 3.4$\mum$ feature polarization
to well exceed what is observed.
These results support the earlier arguments that
the aliphatic hydrocarbon component is physically
separated from the silicate component
unless the 3.4$\mum$ absorption feature is just
a minor carbon sink in the ISM.
%
%
\end{abstract}



\section{Introduction\label{sec:intro}}
Although the exact nature of interstellar dust 
remains uncertain, it is now well recognized 
that interstellar grains consist of amorphous silicates 
and some form of carbonaceous materials.
While the identification of amorphous silicate dust
in the interstellar medium (ISM) is relatively
secure through the broad featureless 
9.7$\mum$ Si--O stretching and 
18$\mum$ O--Si--O bending absorption features
(see Henning 2010),
our understanding of the carbon dust component,
mainly through the 2175$\Angstrom$ extinction bump 
and the 3.4$\mum$ C--H absorption feature
is not as clear. 

All contemporary dust models assume carbon dust 
as a key grain component. They differ mainly
in terms of the exact chemical and physical forms
the carbon dust component take:
(1) The silicate-graphite model 
    (Mathis et al.\ 1977; Draine \& Lee 1984; 
     Siebenmorgen \& Kr\"ugel 1992;
     Weingartner \& Draine 2001; Draine \& Li 2007)
    assume that graphite is the major carbon sink
    and the silicate and graphite components 
    are bare and physically separated;
(2) The silicate core-carbonaceous mantle model
      (D\'{e}sert et al.\ 1990; 
       Jones et al.\ 1990; Li \& Greenberg 1997; 
       Jones et al.\ 2013)
      assumes that silicate grains are coated
      with a carbonaceous mantle made of 
      either hydrogenated amorphous carbon (HAC)
      or organic refractory;
(3) The composite model (Mathis \& Whiffen 1989; 
    Mathis 1996; Zubko et al.\ 2004)
    assumes the dust to be low-density aggregates of 
    small silicates and carbonaceous particles
    (amorphous carbon, HAC, and organic refractories).
All dust models appear to be in general agreement 
with the observational constraints,
including the interstellar extinction,
scattering, polarization, IR emission 
and interstellar depletion. 
  
More recently, spectropolarimetry of the 3.4$\mum$ interstellar
absorption feature has been used to distinguish between dust models
(Adamson et al.\ 1999; Ishii et al.\ 2002; 
Chiar et al.\ 2006; Mason et al.\ 2007).
The 3.4$\mum$ absorption feature, commonly attributed to 
the C--H stretching mode in saturated aliphatic hydrocarbon dust
(see Pendleton \& Allamandola 2002),
is ubiquitously seen in the diffuse ISM
of the Milky Way and external galaxies
(e.g., see Mason et al.\ 2004).\footnote{%
  Chiar et al.\ (2013) argued that the carrier of 
  the 3.4$\mum$ feature is largely aromatic 
  instead of aliphatic.
  But see Dartois et al.\ (2007)
  who argued that the carrier of 
  the 3.4$\mum$ feature is highly aliphatic.
  }

The silicate core-carbonaceous mantle model assumes 
that the 3.4$\mum$ absorption feature arises in 
the hydrocarbon mantles coating the amorphous silicate
cores (Jones et al.\ 1990; Li \& Greenberg 1997). 
The hydrocarbon mantles consist of either
``organic refractory'' (Greenberg et al.\ 1995)
or HAC  (Jones et al.\ 1990). 
Both ``organic refractory'' and HAC provide a close
  match to the interstellar 3.4$\mum$ absorption feature
  (Greenberg et al.\ 1995; Mennella et al.\ 1999). 
  The interstellar organic refractory material 
  is essentially HAC in character.
  The major difference between the organic refractory material
  with HAC lies in the way how they are made: 
  the former is derived from the UV photo-processing 
  of interstellar ice mixtures 
  accreted on the pre-existing silicate cores
  (Greenberg et al.\ 1995); 
  the latter results from direct accretion of 
  gas-phase elemental carbon on the silicate cores 
  in the diffuse ISM (Duley et al.\ 1989;
  Jones et al.\ 1990).

The 9.7$\mum$ and 18$\mum$ silicate absorption features 
have been reported to be polarized along various sightlines 
probing both the diffuse ISM and dense molecular clouds
(see Smith et al.\ 2000, Wright et al.\ 2002, Aitken 2005),
suggesting that the silicate component is nonspherical and aligned.
If the carrier of the 3.4$\mum$ feature resides
in the carbonaceous mantles on the silicate cores,
we would expect the 3.4$\mum$ absorption feature 
to be polarized as well (see Li \& Greenberg 2002).

However, all spectropolarimetric observations 
show that the 3.4$\mum$ absorption feature is 
essentially unpolarized (Adamson et al.\ 1999, Ishii et al.\ 2002,
Chiar et al.\ 2006, Mason et al.\ 2007). 
This, especially the nondetection of the 3.4$\mum$ feature polarization
in the Galactic center Quintuplet combined with the fact
that the 9.7$\mum$ silicate feature is polarized 
in the same sightline (Chiar et al.\ 2006), 
poses a severe challenge against the core-mantle model.
They argue that the hydrocarbon dust do not reside on 
the same grains as the silicates,
and likely form a separate population 
of small grains which are either spherical 
or not sufficiently aligned. 

This challenge might be alleviated if the mantle
is much less elongated than the silicate core
so that the 3.4$\mum$ feature would be polarized 
to a much smaller degree than the 9.7$\mum$ silicate feature. 
The aim of this work is to investigate  this secanario.
We consider an extreme case:
the dust consists of a {\it spheroidal} silicate core
coated with a layer of {\it spherical} aliphatic 
hydrocarbon mantle (\S2). 
To be complete, we also consider 
the composite model in \S3.
We discuss these results in \S4.



\vspace{-5mm}
\section{Spheroidal Core-Spherical Mantle Dust
         \label{sec:mod}}
Interstellar polarization is caused by the differential
extinction of the 2 perpendicular electric vectors of
starlight by aligned, nonspherical grains.
The less elongated the carrier of the 3.4$\mum$ feature is, 
the less is the degree to which the 3.4$\mum$ feature will be polarized. 
Therefore, a lower limit on the 3.4$\mum$ absorption polarization 
will be achieved if the hydrocarbon mantles are {\it spherical}, 
while the silicate cores are {\it elongated}.
We therefore consider 
{\it spheroidal} silicate core-{\it spherical} carbonaceous mantle grains.

Let $a_c$ and $b_c$ be the core semi-axis along and
perpendicular to the symmetry axis respectively; 
$r$ be the spherical radius of the mantle. 
Let $\vsil$ and $\vor$ be the silicate core and 
carbonaceous mantle volumes, respectively.
The mantle-to-core volume ratio:
$V_{\rm{carb}}/V_{\rm{sil}}
=({r^{3}-a_{c}b_{c}^{2}})/{a_{c}b_{c}^{2}}$ can be 
estimated from the observed optical depths
  of the 3.4$\mum$ hydrocarbon feature ($\tauor$)
  and the 9.7$\mum$ silicate feature ($\tausil$):
  $\vor/\vsil \approx \left(\tauor/\tausil\right)
  \left(\rho_{\rm sil}/\rho_{\rm carb}\right)
  \left(\kappa_{\rm sil}^{9.7}/\kappa_{\rm carb}^{3.4}\right)
  \approx 0.25$ 
  where $\tauor/\tausil \approx 1/18$ 
  (Sandford et al.\ 1995); 
  $\rho_{\rm sil}$ and $\rho_{\rm carb}$
  are the mass densities of the silicate ($\approx 3.5\g\cm^{-3}$)
  and carbonaceous dust ($\approx 1.5\g\cm^{-3}$);
  $\kappa_{\rm sil}^{9.7}$
  is the 9.7$\mum$ Si--O silicate mass absorption coefficient
  ($\approx 2850\cm^2\g^{-1}$; Draine \& Lee 1984);
  $\kappa_{\rm carb}^{3.4}$ 
  is the 3.4$\mum$ C--H mass absorption coefficient of 
  carbonaceous organic refractory dust 
  ($\approx 1500\cm^2\g^{-1}$; see Li \& Greenberg 2002).
  However, a much thicker mantle ($\vor/\vsil \approx 1$)
  is required to account for the visual/near-IR interstellar 
  extinction (Li \& Greenberg 1997). 
  The mass ratio of carbonaceous organics to silicates 
  in the coma of comet Halley, measured {\it in situ}, 
  was approximately 0.5 (Kissel \& Krueger 1987),
  pointing to $\vor/\vsil \approx 1$. 
  It is often suggested that cometary dust is made of 
  interstellar grain aggregates (Greenberg \& Li 1999, 
  Kimura et al.\ 2003, Kolokolova \& Kimura 2010).
  In dense clouds we would expect a thicker hydrocarbon mantle
  (although the 3.4$\mum$ feature is not seen in dense
   molecular clouds; see Mennella et al.\ 2001, Mennella 2010).
  Therefore, we will consider 3 mantle thicknesses: 
  $\vor/\vsil = 0.25, 1, 2$. 
  If the available Si elements (say, $\approx 32\ppm$ 
  per H atom like that of Sun where ppm refers to parts per million; 
  Asplund et al.\ 2009)
  are all depleted in the silicate cores,   
  the carbonaceous mantles require
  C/H$\approx 49$, 196, 392$\ppm$. 
  Given the interstellar abundance constraints
  (e.g. see Li 2005), $\vor/\vsil <1$ seems more reasonable.    
  We consider a wide range of 
  core-elongations of $a_c/b_c$
  which are required to satisfy the constraint of
  $\vsil/(\vor+\vsil) < a_c/b_c < [(\vor+\vsil)/\vsil]^{1/2}$
  since both $a_c$ and $b_c$ must be smaller than $r$.
 
Let $\cpar$ and $\cper$ be the absorption cross sections 
for light polarized parallel and perpendicular, respectively, 
to the grain symmetry axis. For an ensemble of grains spinning
and precessing about the magnetic field, the polarization
cross section is $\cppol = \left(\cpar-\cper\right)/2$
for prolates, and $\copol = \left(\cper-\cpar\right)$ for oblates;
the absorption cross section is   
$\cabs = \left(\cpar+2\cper\right)/3 - 
\Phi \cpol \left(3-2/\cos^2\gamma\right)/6$
where $\Phi$ is the polarization reduction factor;
$\gamma$ is the angle between the magnetic field
and the plane of the sky (Lee \& Draine 1985). 
We take $\Phi=1$ and $\gamma=0$.

\begin{figure}
\centering
\includegraphics[angle=0,width=11cm]{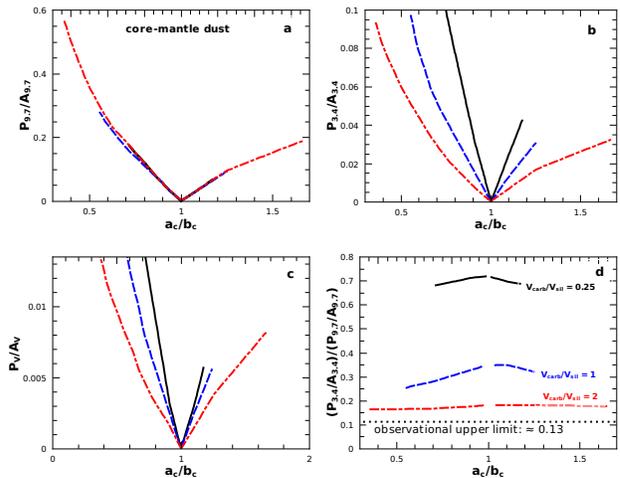}
\vspace{-2.2cm}
\caption{
        \label{fig:P2A}
         Polarization-to-extinction ratios
         of (a) the 9.7$\mum$ silicate feature $\Psil/\Asil$,
         (b) the 3.4$\mum$ hydrocarbon feature $\Pch/\Ach$,
         and (c) the visible band $P_V/A_V$,
         as well as (d) the relative polarization $\PTOA$,
         predicted from the spheroidal silicate core-spherical
         hydrocarbon mantle model as a function
         of core-elongations ($a_c/b_c$) for
         3 mantle thicknesses 
         $\vor/\vsil=0.25$ (solid), 1 (dashed),
         and 2 (dot-dashed).
         In panel (d) the lines break at $a_c/b_c=1$
         since $\Psil/\Asil=0$ and $\Pch/\Ach=0$ for dust
         with $a_c/b_c=1$ (i.e., spherical dust does not cause
         polarization). Also shown in panel (d) is
         the upper limit of $\PTOA \approx 0.13$ (dotted)    
         observed for the 3.4$\mum$ feature polarization of 
         the Galactic center Quintuplet object GCS\,3-II
         (Chiar et al.\ 2006).
         }
\vspace{-5mm}
\end{figure}

For a given core-elongation of $a_c/b_c$ and a given
mantle-to-core volume ratio of $\vor/\vsil$, we use
the discrete dipole approximation of Draine (1988; DDSCAT) 
to calculate the 3.4$\mum$ C--H excess extinction
$\Ach$ and excess polarization $\Pch$, 
as well as the 9.7$\mum$ Si--O excess extinction $\Asil$ 
and excess polarization $\Psil$.
Here by ``excess'' we mean the extinction 
and polarization of the absorption feature in excess
of the continuum extinction and polarization underneath
the feature.
In Figure~\ref{fig:P2A} we show 
the polarization-to-extinction ratio 
as a function of core-elongation $a_c/b_c$ 
for 3 different mantle thicknesses 
$\vor/\vsil=0.25,1,2$ 
for the 9.7$\mum$ silicate feature,
the 3.4$\mum$ hydrocarbon feature,
and the visible band ($P_V/A_V$). 
Also shown in Figure \ref{fig:P2A}
is $\PTOA$, which measures the degree 
to which the 3.4$\mum$ hydrocarbon feature
is polarized relative to the 9.7$\mum$ silicate feature.

As expected, the more elongated the core is, 
the more polarized is the 9.7$\mum$ silicate feature
(see Figure \ref{fig:P2A}a).
This is also true for the 3.4$\mum$ hydrocarbon feature
(see Figure \ref{fig:P2A}b) and the visible band
(see Figure \ref{fig:P2A}c). This can be understood
from the grain geometry: for grains with a prolate 
(oblate) core, the light polarized along the semi-minor 
(semi-major) axis will see more hydrocarbon dust.
Therefore, the more elongated the core is, 
the larger is the difference between the extinction 
of starlight polarized along the semi-major axis
and that along the semi-minor axis.

The mantle thickness has little effect on 
the silicate feature polarization
(see Figure \ref{fig:P2A}a).
But for the 3.4$\mum$ hydrocarbon feature,
it becomes less polarized when the mantle 
becomes thicker (see Figure \ref{fig:P2A}b).
This is not unexpected -- when the mantle 
becomes thicker, the difference
between the amounts of hydrocarbon dust
seen by the two electric vectors of starlight 
becomes smaller. If the spherical mantle is very thick
(i.e. $\vor \gg \vsil$), the 3.4$\mum$ feature
will become essentially unpolarized.

As expected from a combination of 
Figure \ref{fig:P2A}a and Figure \ref{fig:P2A}b,
the relative polarization $\PTOA$ 
decreases as the silicate cores are
more elongated (see Figure \ref{fig:P2A}d);
it also decreases as the hydrocarbon mantle 
becomes thicker. 

\vspace{-5mm}
\section{Composite Model
         \label{sec:composite}}
We now consider the composite dust model in which
interstellar grains are taken to be
fluffy aggregates of small silicates, 
vacuum, and carbon of various kinds 
(amorphous carbon, HAC, and organic refractories; 
Mathis \& Whiffen 1989, Mathis 1996).
We test this model using DDSCAT
and assuming the composite grains are spheroidal-shaped. 

Following Mathis (1996), we take the volume-filling factor of 
vacuum $f_{\rm vac}=0.45$, 
and the mass ratio between hydrocarbon dust and silicate dust
$m_{\rm carb}/m_{\rm sil}=0.7$.
The volume-filling factor is $f_{\rm sil}\approx 0.21$ 
and $f_{\rm carb}\approx 0.34$ for the silicate component
and the hydrocarbon component, respectively
(i.e., $V_{\rm carb}/V_{\rm sil}\approx 1.62$).

Let $\kappa_{\rm abs}(3.4)$ and 
$\kappa_{\rm abs}(9.7)$ respectively be 
the mass-absorption coefficient of 
the 3.4$\mum$ feature 
and the 9.7$\mum$ feature 
in excess of the continuum.
For composite grains, we use DDSCAT to
calculate $\kappa_{\rm abs}(3.4)$ and 
$\kappa_{\rm abs}(9.7)$ and derive
$\kappa_{\rm abs}(3.4)\approx 240\cm^{2}\g^{-1}$ and 
$\kappa_{\rm abs}(9.7)\approx 1500\cm^{2}\g^{-1}$,
independent of elongation.

\begin{figure}
\centering
\includegraphics[angle=0,width=8cm]{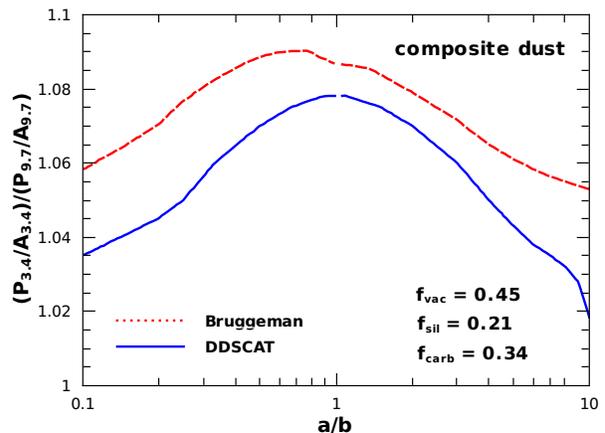}
\vspace{-0.5cm}
\caption{
         \label{fig:P2A2nd}
               Relative polarization $\PTOA$
               predicted from the composite model
               (with $V_{\rm carb}/V_{\rm sil}\approx 1.62$) 
               as a function of elongations of $a/b$
               in the range of $0.1<a/b<10$
               (except $a/b=1$ for which 
                $\Psil/\Asil=0$ and $\Pch/\Ach=0$).
           }
\vspace{-5mm}
\end{figure}

We calculate the polarization-to-extinction ratio 
as a function of elongation $a/b=0.1-10$.
As shown in Figure \ref{fig:P2A2nd}, 
$\PTOA \approx 1.02 - 1.08$. 
We have also calculated the polarization-to-extinction
ratio for composite dust 
using the Bruggeman effective medium theory
(see Bohren \& Huffman 1983).
The results are very close to that calculated from DDSCAT,
only differing by $<$\,2\%
(see Figure \ref{fig:P2A2nd}).
We conclude that the composite model 
predicts a similar degree of polarization 
(i.e., $\Psil/\Asil \approx \Pch/\Ach$) for 
the 9.7$\mum$ Si--O feature and the 3.4$\mum$ C--H feature,
implying that the composite model is also inconsistent with
the nondetection of 3.4$\mum$ C--H polarization feature.

\vspace{-5mm}
\section{Discussion
         \label{sec:discussion}}
Since the spheroidal core-spherical mantle grains 
considered in \S2 represent an extreme 
case in which the 3.4$\mum$ hydrocarbon 
feature is least polarized (relative to the 9.7$\mum$
silicate feature), if in an astronomical object
the 3.4$\mum$ absorption feature is observed to 
have an even lower degree of polarization, 
the core-mantle model will be severely challenged. 

Chiar et al.\ (2006) placed an upper
limit on the 3.4$\mum$ feature polarization of 
the Galactic center Quintuplet object GCS\,3-II:
$\PTOA \approx 0.13$. This upper limit is even
lower than the lower limits predicted from
the spheroidal core-spherical mantle model
with $\vor/\vsil=2$ for which $\PTOA > 0.16$
(see Figure \ref{fig:P2A}d) over the entire
allowable ranges of core-elongations 
($1/3 < a_c/b_c < \sqrt{3}$; see \S2).
This supports the idea put forward by 
Adamson et al.\ (1999) and Chiar et al.\ (2006) that,
based on the nondetection of the 3.4$\mum$ feature polarization,
the core-mantle model is invalid or the carrier 
of the 3.4$\mum$ feature does not reside in the carbonaceous mantle 
as previously thought. The hydrocarbon dust component
responsible for the 3.4$\mum$ feature must be physically 
separated from the silicate component.
This component must be either spherical 
or poorly aligned (or both) so that the resulting
3.4$\mum$ absorption feature is essentially unpolarized.

With a {\it thicker} carbon mantle, one expects a {\it smaller}
3.4$\mum$ polarization-to-extinction ratio $\Pch/\Ach$.
With $\vor/\vsil>3$,  we obtain
$\PTOA < 0.13$ which appears to satisfy 
the observational upper limit of Chiar et al.\ (2006).
However, grains with such a thicker spherical mantle
would produce little polarization 
in the optical wavelength range.
This is inconsistent with the observations
of the interstellar polarization: 
(1) light reaching us from reddened stars is
often polarized in the optical;
(2) the interstellar polarization curve
--- the degree of polarization as a function of wavelength ---
rises from the near-IR ($\lambda\sim 2\mum$), 
has a maximum somewhere in the optical 
($\lambda_{\rm max}\approx 0.55\mum$)
and then decreases toward the ultraviolet 
(UV; see Whittet 2003).
To be considered successful,
a grain model should have its bulk, 
submicrometer-sized dust component 
to be non-spherical and sufficiently aligned
to reproduce the observed interstellar 
polarization curve (e.g., see Voshchinnikov 2012).
The silicate-graphite model requires
either silicate (e.g., Mathis et al.\ 1977, 
Kim \& Martin 1994) or both silicate and 
graphite (see Draine \& Fraisse 2009,
Siebenmorgen et al.\ 2014)
to account for the observed optical polarization.
The core-mantle model requires the core-mantle dust
to produce the interstellar optical polarization
(see Li \& Greenberg 1997).
The composite model requires the porous composite dust
to account for the observed optical polarization
(see Mathis \& Whiffen 1989, Mathis 1998).

We note that, as shown in Figure \ref{fig:P2A}c,
with $\vor/\vsil > 1$, 
the optical polarization-to-extinction ratio 
($P_V/A_V$) predicted from 
the spheroidal core-spherical mantle model
is too small to compare with 
the observational value of
$P_V/A_V \le 0.064$ (Whittet 2003)
which should be achieved for perfectly aligned grains. 
This indicates that, although with a thick spherical
carbon mantle one may satisfy the observed
upper limit of $\PTOA<0.13$,
the starlight will essentially see the dust as
{\it spherical} and will not be polarized 
in the optical.

The composite model is not able to alleviate 
the 3.4$\mum$ polarization challenge.
With $\PTOA \approx 1.02 - 1.08$
(compared with the observed upper limit of
$\PTOA\approx 0.13$; Chiar et al.\ 2006),
the composite model predicts a similar degree 
of polarization for the 9.7$\mum$ Si--O feature 
and the 3.4$\mum$ C--H feature.    
This indicates that the 3.4$\mum$ hydrocarbon feature 
should have a positive detection for the lines of sight
along which the 9.7$\mum$ silicate feature is observed
to be polarized. The nondetection of the 3.4$\mum$ feature 
polarization in the Galactic center Quintuplet combined with 
the detection of the 9.7$\mum$ silicate feature polarization 
in the same sightline (Chiar et al.\ 2006) poses a severe 
challenge against the composite model.

The core-mantle model may remain valid 
if the mantle component does not contain
the carrier of the 3.4$\mum$ absorption feature,
i.e., the carrier of the 3.4$\mum$ absorption 
feature is not a major carbon sink in the ISM
and is physically not associated with 
the bulk core-mantle dust. 
Jones et al.\ (2013) argued that the aliphatic hydrocarbon
material is subject to UV photo-processing in the diffuse ISM
and is expected to be maximally-aromatized in the order of 
a million years. Therefore, they suggested
that the mantle material of the core-mantle dust is mainly 
aromatic and is not responsible for 
the 3.4$\mum$ absorption feature.
According to Jones et al.\ (2013), 
the 3.4$\mum$ absorption feature is 
due to a separate population of 
small aliphatic hydrocarbon dust.

If the core-mantle dust is not responsible for
the 3.4$\mum$ absorption feature, 
we might encounter a carbon budget problem: 
if the total carbon abundance (relative to H)
in the ISM is like the Sun $\cism=\csun\approx 224\ppm$
(Asplund et al.\ 2009)\footnote{%
   We note that the C abundance of the early B stars 
   which are thought to be ideal indicators for 
   the present-day interstellar abundances 
   since they preserve their pristine abundances
   is close to the solar C abundance:
   $\cBstar\approx 214\pm20\ppm$ (Przybilla et al.\ 2008)
   and $\cBstar\approx 209\pm15\ppm$ (Nieva \& Przybilla 2012).
   }
or proto-Sun $\cism=\csun\approx 288\ppm$
(Lodders 2003),
with $\cgas\approx 140\ppm$ in the gas phase
(Cardelli et al.\ 1996)   
and $\cpah\approx 60\ppm$ in PAHs (Li \& Draine 2001)
subtracted, there is only $\cdust\approx 24\ppm$
or $\cdust\approx 57\ppm$
left for the 2175$\Angstrom$ extinction bump,
the 3.4$\mum$ absorption feature, 
the ``extended red emission'' (ERE) 
which is most likely from some sort of 
small carbon-based dust (Witt \& Vijh 2004),
and a population of bulk carbon dust.\footnote{%
   Sofia et al.\ (2011) derived 
   $\cgas\approx 100 \ppm$ for several interstellar
   sightlines from the strong transition of C\,II] 
   at 1334$\Angstrom$. They argued that the oscillator 
   strength for the C\,II] transition at 2325$\Angstrom$ 
   previously used by Cardelli et al.\ (1996)
   to obtain $\cgas\approx 140\ppm$ might have been 
   underestimated.
   But even with $\cgas\approx 100 \ppm$, 
   the amount of C available for the 2175$\Angstrom$ bump,
   the 3.4$\mum$ feature, the ERE, and the bulk carbon dust,
   is only $\cdust\approx 64\ppm$
   or $\cdust\approx 97\ppm$
   and does not seem to be sufficient.
   }
The latter is required to account for part of the visual
extinction since silicates alone are not able to provide 
enough extinction (see Footnote-14 in Li 2004).

Furthermore, according to Draine (1990)
and Jones et al.\ (1994), 
most of the dust mass in the ISM was condensed in the ISM, 
it is not very clear how it is possible for
the re-condensation to keep the silicate 
and carbon grain populations apart in the ISM.
Draine (2009) postulated a scenario 
for the ISM to grow two distinct grain types 
(i.e., silicate and carbon dust) out of a single 
gas mixture. He argued that when Mg, Si, Fe, and 
O atoms and ions arrive at the amorphous silicate
surface, they are able to grow additional amorphous 
silicate; in contrast, the C atom physisorbed 
on the amorphous silicate surface 
might undergo photoexcitation to an excited state 
that is repulsive, ejecting it from the surface. 
Or perhaps the C would become hydrogenated or oxidized,
with the resulting CH or CO undergoing photodesorption 
from the surface. Such processes could keep the amorphous 
silicate carbon-free in the diffuse ISM.
Similar processes may occur on exposed carbonaceous 
surfaces: impinging C atoms could grow new carbonaceous 
material, whereas impinging Mg, Si, Fe atoms
could be removed by some combination of reaction 
with impinging H or O, and photoexcitation by UV.

\vspace{-5mm}
\section*{Acknowledgements}
We thank A.P.~Jones, A.~Mishra, N.V.~Voshchinnikov, 
and the anonymous referee for helpful suggestions. 
We thank B.T.~Draine for making 
the DDSCAT code available. 
We are supported in part by
NSF AST-1109039, NNX13AE63G, NSFC\,11173019, 
and the University of Missouri Research Board.

\bsp
\label{lastpage}

\end{document}